# Temperature dependent magnetic, dielectric and Raman studies of partially disordered $La_2NiMnO_6$

Pradeep Kumar[1], S. Ghara[2], B. Rajeswaran[2], D. V. S. Muthu[1], A. Sundaresan[2], and A. K. Sood[1*]

[1]Department of Physics, Indian Institute of Science, Bangalore -560012, India

[2]Chemistry and Physics of Materials Unit, Jawaharlal Nehru Centre for Advanced Scientific Research,

Bangalore-560064, India

**ABSTRACT**

We report a detailed magnetic, dielectric and Raman studies on partially disordered and biphasic double perovskite $La_2NiMnO_6$. DC and AC magnetic susceptibility measurements show two magnetic anomalies at $T_{C1}$ ~ 270 K and $T_{C2}$ ~ 240 K, which may indicate the ferromagnetic ordering of the monoclinic and rhombohedral phases, respectively. A broad peak at a lower temperature ($T_{sg}$ ~ 70 K) is also observed indicating a spin-glass transition due to partial anti-site disorder of $Ni^{2+}$ and $Mn^{4+}$ ions. Unlike the pure monoclinic phase, the biphasic compound exhibits a broad but a clear dielectric anomaly around 270 K which is a signature of magneto-dielectric effect. Temperature-dependent Raman studies between the temperature range 12 K to 300 K in a wide spectral range from 220 $cm^{-1}$ to 1530 $cm^{-1}$ reveal a strong renormalization of the first as well as second-order Raman modes associated with the $(Ni/Mn)O_6$ octahedra near $T_{C1}$ implying a strong spin-phonon coupling. In addition, an anomaly is seen in the vicinity of spin-glass transition temperature in the temperature dependence of the frequency of the anti-symmetric stretching vibration of the octahedra.

# 1. INTRODUCTION

There is a recent upsurge of interest in the double-perovskite, $La_2NiMnO_6$ (LNMO) due to its rich physics, particularly large dielectric anomaly and the related potential applications [1-6]. Much effort has recently been devoted to a better understanding of the coupling between the magnetic, phononic and electronic degrees of freedom (DOF) because an intricate interplay between these DOF is believed to be responsible for the various novel physical phenomena observed in this system [1]. Magnetization and neutron diffraction studies have established LNMO to be a ferromagnet at near room temperature ($T_C \sim 270$ K) [7]. It has a monoclinic ($P2_1/n$) structure at ambient temperature and a rhombohedral ($R\text{-}3$) structure at high temperatures [3,8]. Often, these two phases coexist at room temperature with the majority phase being the monoclinic. However, one can obtain a pure monoclinic phase under stringent synthesis conditions which not only determines the formation of these phases but also control the formation of lanthanum and oxygen vacancies and anti-site disorder [9-11]. Presence of dielectric anomaly and a large magneto-capacitance (MC) effect has been reported in the temperature interval, 220 K - 280 K near magnetic transition temperature ($T_c \sim 280$ K) in the highly ordered but biphasic LNMO system [1]. A multiglass behavior and MC effect over a wide range of temperature (150 K - 300 K) have been observed in the partially disordered single phase monoclinic system where the latter has been explained to be due to asymmetric hopping of electrons between $Ni^{2+}$ and $Mn^{4+}$ ions. It is to be noted that dielectric studies on pure rhombohedral phase of LNMO is not known. Meanwhile, an extrinsic magneto-dielectric effect has been reported in nanoparticles of monoclinic LNMO [12]. Further, anti-site disorder between $Ni^{2+}$ and $Mn^{4+}$ ions leads to $Mn^{4+}\text{-}O^{2-}\text{-}Mn^{4+}$ and $Ni^{2+}\text{-}O^{2-}\text{-}Ni^{2+}$ antiferromagnetic coupling resulting in a spin-glass behavior in the vicinity of 70 K ($T_{sg}$). However, the dominant $Ni^{2+}\text{-}O^{2-}\text{-}$ $Mn^{4+}$ super-exchange coupling of the ordered structure remains ferromagnetic below the transition temperature ( ~ 270 K).

The coupling of soft infrared phonons with Ni and Mn spins via super-exchange interactions has been shown to occur in the rhombohedral phase using first-principles calculations [13]. These soft phonon modes were suggested to be responsible for the observed dielectric anomaly as well as the MD effect near $T_c$ in the multiphase sample [1]. It should be noted that no dielectric anomaly has been reported near $T_c$ in a pure monoclinic phase [6,9-10,12,14]. Since there are no calculations for the monoclinic phase and the fact that the dielectric anomaly is observed only in a multiphase sample, it become necessary to understand the origin of dielectric anomaly which may be related to rhombohedral phase or the multiphase nature of the sample.

A significant magneto-dielectric effect in double-perovskite systems imply a strong coupling between the lattice and the magnetic degrees of freedom [5, 15]. It therefore, becomes important to study the role of phonons as a function of temperature using Raman spectroscopy which is a well proven powerful technique to investigate the spin-phonon coupling, charge/orbital ordering, long-range cation ordering and other structural changes in perovskite oxides [16-20]. Main focus of the already reported studies in the literature has been the first-order Raman modes [2, 4, 21, 22-26] as a function of temperature showing the signatures of spin-phonon coupling; in particular a phonon mode associated with the symmetric stretching vibration of $(Ni/Mn)O_6$ octahedra shows continuous softening down to the lowest temperature in the ferromagnetic phase.

Here, we report magnetic, dielectric and Raman studies on partial disordered and biphasic LNMO bulk in the temperature range 12 K to 300 K. We observe a broad dielectric anomaly centered around 270 K. The observed first-order symmetric stretching mode (S2) as well its overtone mode (S5) show anomalous behaviour due to strong spin-phonon coupling. In addition, the anti-symmetric stretching vibration (S1) of the oxygen octahedra shows an anomaly near a temperature associated with spin-glass transition ~ 70 K, as inferred from magnetic susceptibility measurements. The anomalous temperature dependence of the integrated intensity ratio of second to first-order phonon modes evidence the occurring of conventional resonant Raman scattering.

## 2. EXPERIMENTAL DETAILS

Polycrystalline samples of the composition $La_2NiMnO_6$ were prepared by the standard solid state reaction method by mixing the individual oxides and heating the mixture at high temperatures (900°C). Final heating was carried at 1400°C and the sample was cooled slowly (1°/min) to room temperature. The sample was characterized by *x*-ray diffraction method using the Bruker D8 Discover diffractometer with Cu Kα radiation. Magnetization measurements were carried out using a Superconducting Quantum Interference Device combined with Vibrating Sample (SQUID VSM) and AC susceptibility measurements were performed utilizing the ACMS option in a Physical Property Measuring System (PPMS), Quantum Design, USA. Dielectric measurements were performed using the E4980 Agilent LCR meter using the multifunctional probe utilizing the cryo system of PPMS, with silver paste used as electrodes. Raman spectroscopic measurements were performed at low temperatures in backscattering geometry, using 514.5 nm line of an Ar-ion Laser (Coherent Innova 300) with Laser power ~ 4 mW at sample. Temperature scanning from 12 K to 300 K was done using a close cycle cryostat with a temperature accuracy of ± 0.1 K. The scattered light was analyzed using a Raman spectrometer (DILOR XY) coupled to liquid nitrogen cooled CCD, with an instrumental broadening of ~ 4 $cm^{-1}$.

## 3. RESULTS AND DISCUSSION

Figure 1 shows the Rietveld refinement of *x*-ray diffraction data at room temperature. The inset shows the data in a narrow angle range showing the biphasic nature of the sample, fitted to give the coexistence of both the phases (58% ($P2_1/n$) and 42% ($R-3c$)). Figure 2 shows the dc magnetization (M) as a function of temperature for zero field cooled (ZFC) and field cooled (FC) at a low field (0.01 T). It could be observed from ZFC as well as FC magnetization that the material shows two ferromagnetic transitions around 270 K and 240 K which is consistent with the presence of two phases as revealed by the *x*-ray diffraction studies. As the ferromagnetic transition temperature in the pure monoclinic phase is found to be near 270 K [7], we attribute the ferromagnetic transition at 240 K to the rhombohedral phase. The saturation magnetic moment obtained from M(H) measurement at 2 K (see inset of Fig.2) is 3.571 $\mu_B$/formula unit(f.u) which signifies anti-site disorder of $Ni^{2+}$ and $Mn^{4+}$ ions. To probe the nature of magnetism in more detail and to also bring out the relevance of the two phases and their impact on the magnetism, AC susceptibility ($\chi$) measurements were performed at various frequencies ranging from 50 Hz to 500 Hz as shown in Fig. 3(a,b). Two significant features can be noticed clearly from the real part ($\chi'$) as well as imaginary part ($\chi''$). There is a frequency-dependent anomaly in the vicinity of 60 K in $\chi''$ (see inset of Fig. 3(b) ) and a double peak feature in $\chi''$ near the ferromagnetic transition temperatures 240 K and 270 K. There is a frequency dispersion in $\chi''$ below $T_c$ but the peak positions do not shift in temperature with frequency. This may result from the frustration caused by ferromagnetic interactions arising from $Ni^{2+}$ - O - $Mn^{4+}$ coupling and antiferromagnetic interactions due to anti-site disorder. The frequency-dependent cusp in the low temperature regime can be attributed to the spin-glass behavior as reported earlier [10].

Figure 4 shows the frequency-dependence of the real part of dielectric constant ($\varepsilon_r$) and loss ($\tan\delta$) in the temperature range 200 K - 320 K. Dielectric constant is relatively independent of temperature up to 100 K (not shown here), above which it undergoes a Maxwell-Wagner like relaxation [10]. The important feature to note in the present study is the observation of a borad hump in the real part of dielectric constant as well as an anomaly in the loss in the vicinity of $T_{C1}$ which are not seen in the pure monoclinic samples [10,12]. Moreover, it is quite different from the first report of magnetocapacitance in $La_2NiMnO_6$ where a step like anomaly was reported at 220 K [1]. As there is no dielectric anomaly in the pure monoclinic sample, we suggest that the dielectric anomaly observed in the present study may be related to the biphasic nature of the sample.

Figure 5 shows the Raman spectrum at 12 K displaying five modes labeled as S1 to S5. The spectra are fitted to a sum of Lorentzians functions. Following the previous studies [2, 4, 23] two strong modes S1 (~ 530 $cm^{-1}$) and S2 (~ 668 $cm^{-1}$) are assigned to the anti-symmetric and symmetric stretching vibration of the $(Ni/Mn)O_6$ octahedra, respectively. In addition to the first-order Raman bands, we observe three high frequency bands around 1068 (S3), 1206 (S4) and 1324 (S5) $cm^{-1}$. Similar high frequency modes have also been reported for LNMO thin films but only at room temperature [4], where two broad peaks observed near 1210 and 1345 $cm^{-1}$ were attributed to the multiphonon Raman scattering. Mode S3 and S5 can be assigned as the overtone of modes S1 and S2, respectively, and mode S4 as a combination of mode S1 and S2. As second-order phonon scattering involves the phonons over the entire Brillion-zone, therefore the frequencies of the observed second-order phonons are not necessarily double of the first-order phonons at the $\Gamma$- point. It can be seen from Fig. 5 that the two strong first-order modes observed near 530 $cm^{-1}$ and 668 $cm^{-1}$ are slightly asymmetric, which can be explained following the work of Bull et al [2] as follows. First, the Ni and Mn sites may not be completely ordered, and the different Ni/Mn-O stretching vibrations lie close in frequency, so that the unresolved contributions from different $NiO_6$ and $MnO_6$ environment may present within the band envelope. Second, the domains with different degrees of Ni and Mn order will result in different contributions to the Ni/Mn-O stretching bonds.

In Fig. 6 we have plotted the mode frequency and linewidth for the four prominent modes, i.e. S1, S2, S4 and S5, as a function of temperature. The following observations can be made: (i) The symmetrical stretching mode (S2) shows an anomalous softening below paramagnetic-to-ferromagnetic transition temperature ($T_{C1}$ ~ 270 K) down to 12 K. The solid line is a linear fit with a slope 0.0156 cm$^{-1}$/K. We note that similar softening below $T_{C1}$ has been reported for the thin films and bulk samples of LNMO and LCMO [21, 23, 26-27] attributed to the strong spin-phonon coupling in the ferromagnetic phase in line with the earlier experimental as well theoretical studies on similar systems [17-18, 20, 23, 25, 28-29]. (ii) In comparison, the temperature dependence of the anti-symmetric stretching mode (S1), being reported for the first time is different: the mode frequency increases with decrease in temperature due to anharmonic effects till 70 K, but shows anomalous softening near $T_{sg}$. The solid lines are linear fits in the two temperature ranges. (iii) The linewidth of the modes S1 and S2 shows normal temperature dependence i.e. linewidth decreases with decrease in temperature, however below $T_{sg}$ the linewidth of S2 mode shows a small anomalous increase, coinciding with the onset of spin-glass state. The solid lines in these panels show a fit to a simple model of cubic anharmonic contribution to the linewidth, where a phonon of frequency $\omega$ decay into two phonons of equal frequency [30]: $\Gamma(T) = \Gamma(0) + C[1 + 2n(\omega(0)/2)]$, where $\Gamma(T)$ and $\Gamma(0)$ are the phonon frequencies at temperature T and T = 0 K, C is Self-energy parameter for a given phonon mode and $n(\omega) = 1/[\exp(\hbar\omega/\kappa_B T) - 1]$ is the Bose-Einstein mean occupation factor. The fitted parameters for the modes S1 and S2 are $\Gamma_0^{S1} = 4.5 \pm 0.7$ cm$^{-1}$, $C^{S1} = 9.4 \pm 1.2$ cm$^{-1}$, $\omega_0^{S1} = 534.8 \pm 0.3$ cm$^{-1}$ and $\Gamma_0^{S2} = 6.1 \pm 0.8$ cm$^{-1}$, $C^{S2} = 11.3 \pm 0.9$ cm$^{-1}$, $\omega_0^{S2} = 666.4 \pm 0.5$ cm$^{-1}$, respectively. (iv) The frequencies of the second-order modes S4 (combination of S1 and S2) and S5 (overtone of S2) also show anomalous temperature-dependence, similar to the first-order S2 mode. Solid line for the mode S5 is a linear fit with slope 0.0125 cm$^{-1}$/K. These anomalous temperature-dependence of the modes S2, S4 and S5 confirm clear signature of strong spin-phonon coupling. The temperature dependence of the mode S1 shows signature of a change in magnetic ordering below $T_{sg}$. The observed spin-phonon coupling has been understood as follows [31]: if an ion is displaced from its equilibrium position by "$x$", then the crystal potential is given as $U = \frac{1}{2}(kx^2) + \sum_{ij} J_{ij} S_i S_j$, where k represents the force constant and the second term arises from spin-spin interactions. As a result of such additional perturbation in the potential due to spin-spin interaction, the phonon frequency is affected by the additional term $(\Delta\omega) = \lambda < S_i S_j >$, where $\lambda = (\partial^2 J_{ij}(x)/\partial x^2)$ is the spin-phonon coupling coefficient and $< S_i S_j >$ is the spin-correlation function. The spin-phonon coupling coefficient, $\lambda$, can be positive or negative and can be different for different phonons.

Figure 7 shows the temperature-dependence of the integrated-intensity ratio of the second-order modes S5 and S4 with respect to their first-order counter part mode S2. It can be seen that the intensity ratios increase with the decreasing temperature. We note that the anomalous increases in intensity with decreasing temperature does not follow the normal Raman process, where the intensity ratio should have increased with the increasing temperature due to phonon population. The Franck-Condon mechanism should give a constant intensity ratio [32]. The observed intensity ratio can arise due to resonant Raman scattering [33-35]. The relevant energy levels involved in resonant Raman scattering using a laser photon energy of 2.41 eV are the Ni and Mn *3d* bands, as shown by the recent DFT calculations [13].

## 4. CONCLUSIONS

In conclusion, we observe a dielectric anomaly in partially disordered and biphasic $La_2NiMnO_6$ around 270 K. We also find anomalies in the AC susceptibility near Curie temperature as well as at a lower temperature (~70 K) which has been understood to be due to the emergence of the spin-glass state. The temperature dependent Raman measurements reveal a strong spin-phonon coupling as reflected in the anomalous temperature dependence of the phonon modes. The anomalous increase in the intensity of the second-order phonons with decreasing temperature is attributed to the resonant Raman scattering. Our results obtained here suggest that the interplay between phononic and magnetic degrees of freedom is crucial to understand the underlying physics of double perovskites. Also, to be noted is the fact that such a dielectric anomaly is not seen in single phase LNMO. We suggest that a study of dielectric properties in a single rhombohedral phase of LNMO with different degrees of anti-stie disorder is required to completely comprehend the emergence of such interesting dielectric properties.

## ACKNOWLEDGMENTS

PK acknowledge CSIR, India, for research fellowship. AKS acknowledge the DST, India, for financial support. AS would like to thank C. Madhu for making the sample.

**FIGURE CAPTIONS:**

**Figure 1**: (Colour online) Rietveld refinement on *x*-ray diffraction data of partially disordered polycrystalline $La_2NiMnO_6$. Inset shows the biphasic nature of the sample. First and second row tick marks are the allowed reflection for monoclinic and rhombohedral phases, respectively.

**Figure 2**: (Colour online) Zero-Field-Cooled (ZFC) and Field-Cooled (FC) magnetization as a function of temperature in $La_2NiMnO_6$. Inset shows the $\mu_B$/(f.u); the presence of anti-site disorder can be inferred from the same.

**Figure 3**: (Colour online) (a) Real and (b) Imaginary part of AC magnetic susceptibility as a function of temperature at various frequencies. Inset in (b) shows the frequency dependenc of the peak around 60 K in a narrow temperature range.

**Figure 4**: (Colour online) Frequency dependence of (a) dielectric constant and (b) loss tangent as a function of temperature.

**Figure 5**: (Colour online) Raman spectra of $La_2NiMnO_6$ measured at 12 K. Thick solid line shows the total fit, thin solid lines show the individual Lorentzian fit.

**Figure 6**: (Colour online) Temperature dependence of the modes S1, S2, S4 and S5. The solid lines in left panel for the modes S1, S2 and S5 are linear fits. Solid lines in right panel for the mode S1 and S2 are fitted curve as described in the text.

**Figure 7**: Temperature dependence of the integrated-intensity ratio of the mode S5 to S2 and S4 to S2.

**Figure 1:**

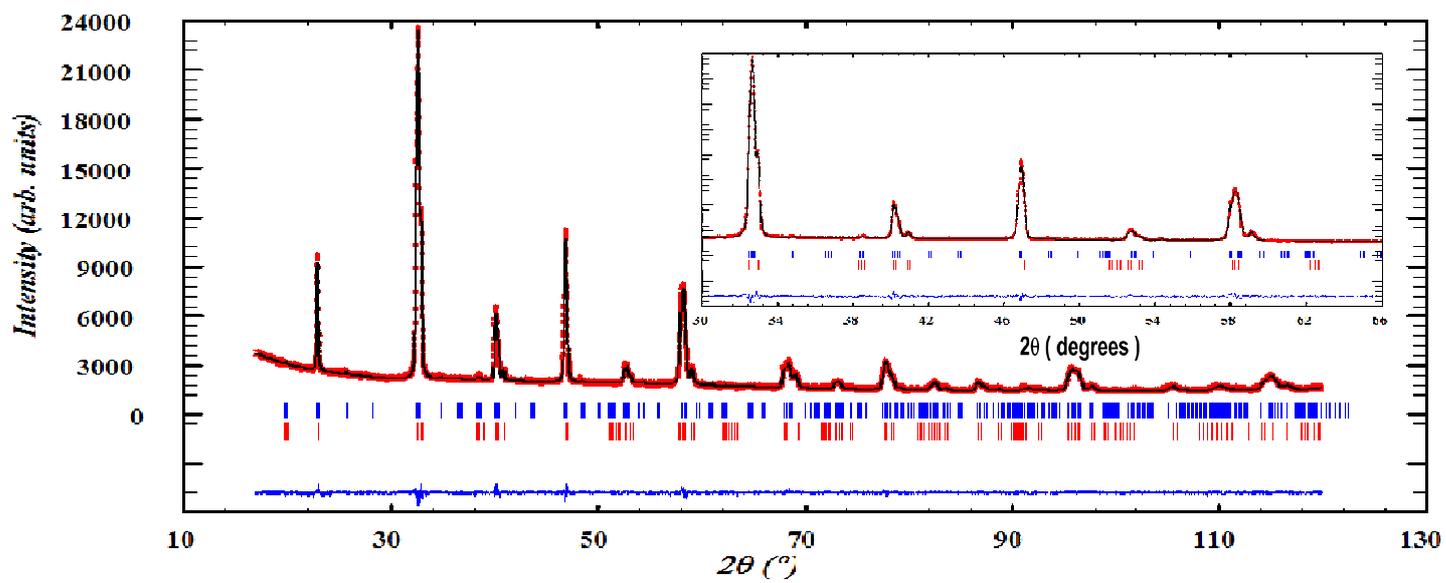

**Figure 2:**

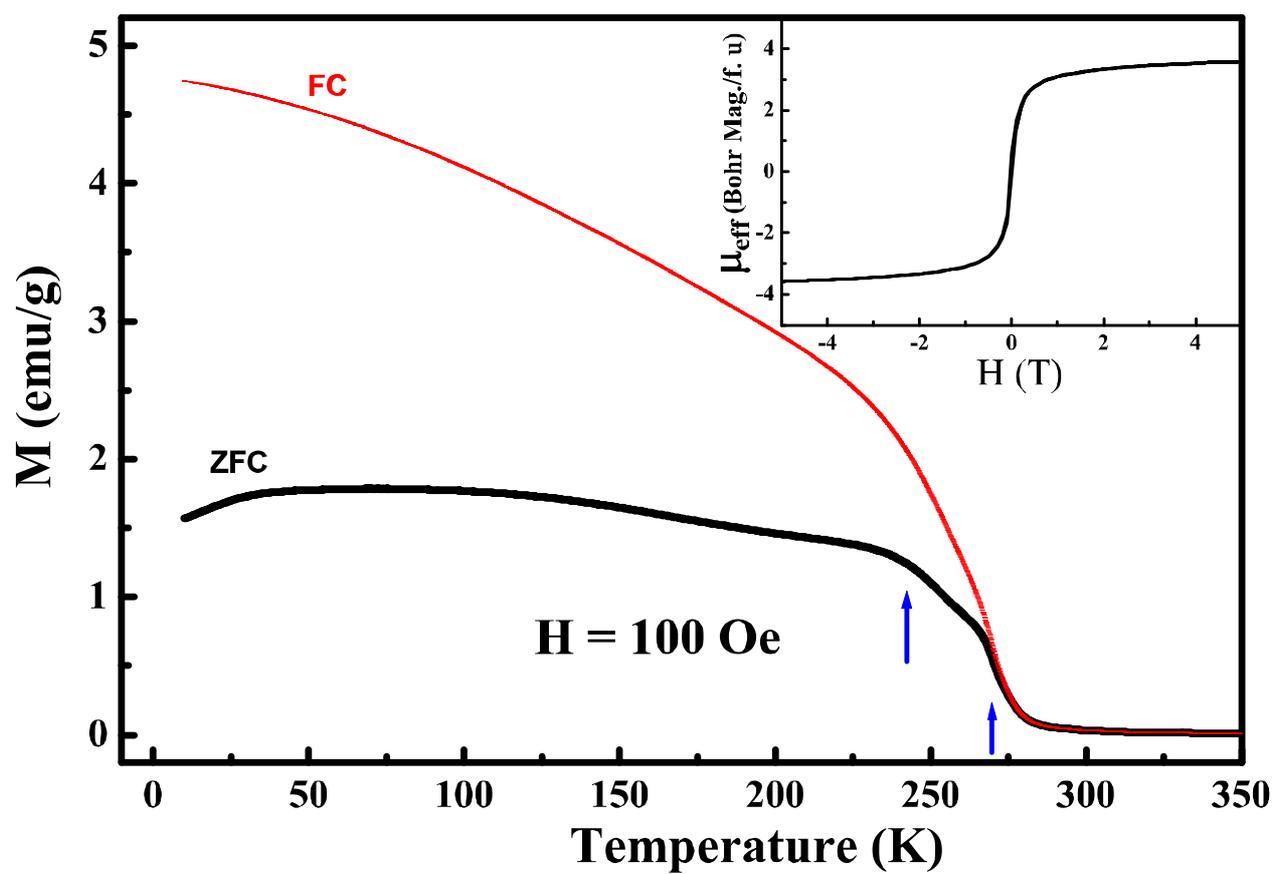

**Figure 3:**

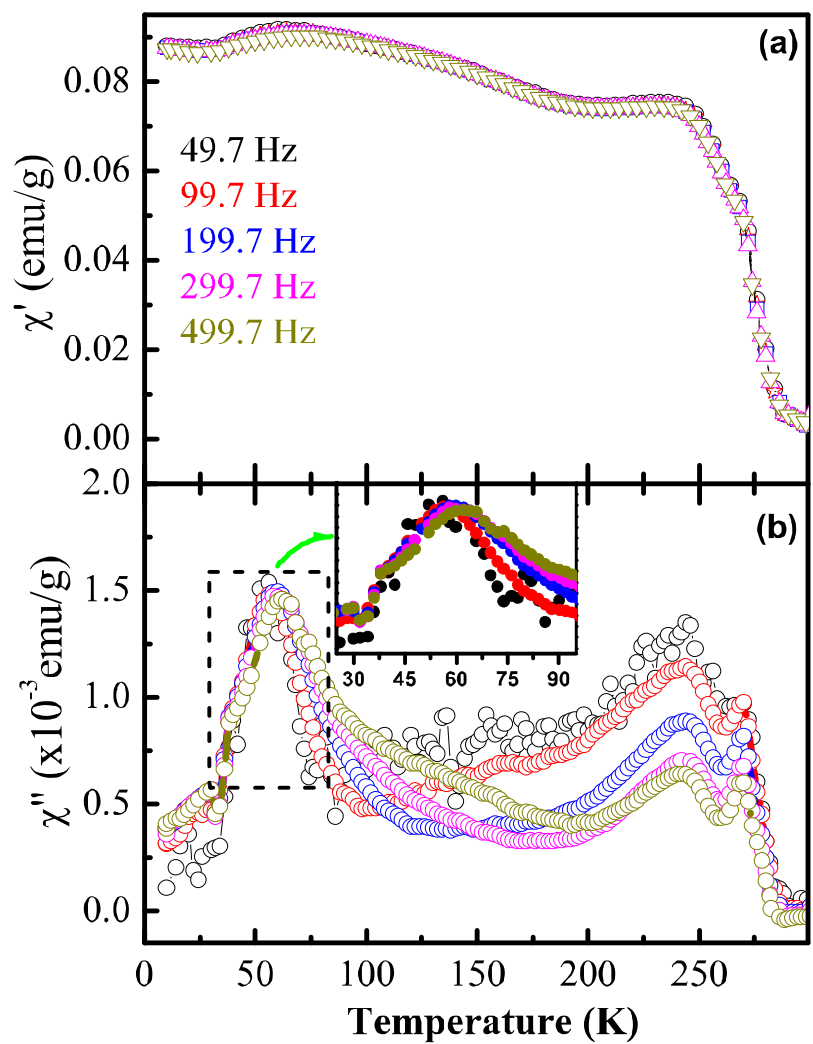

**Figure 4 :**

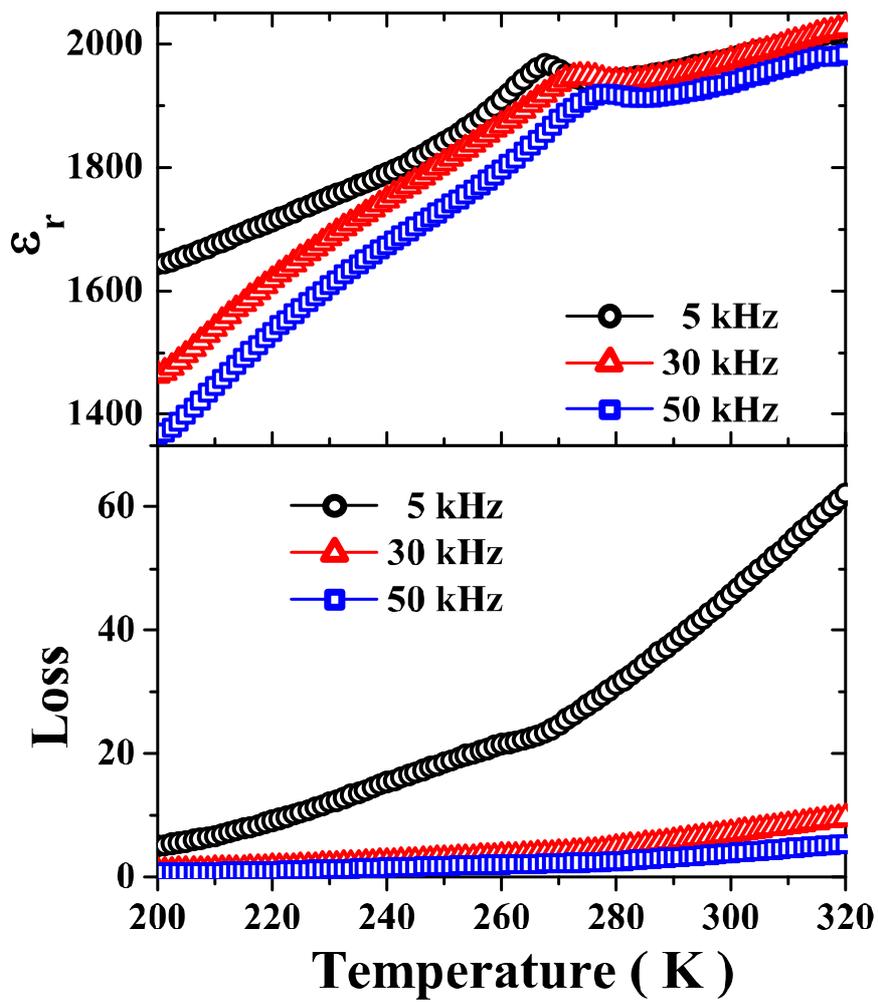

**Figure 5:**

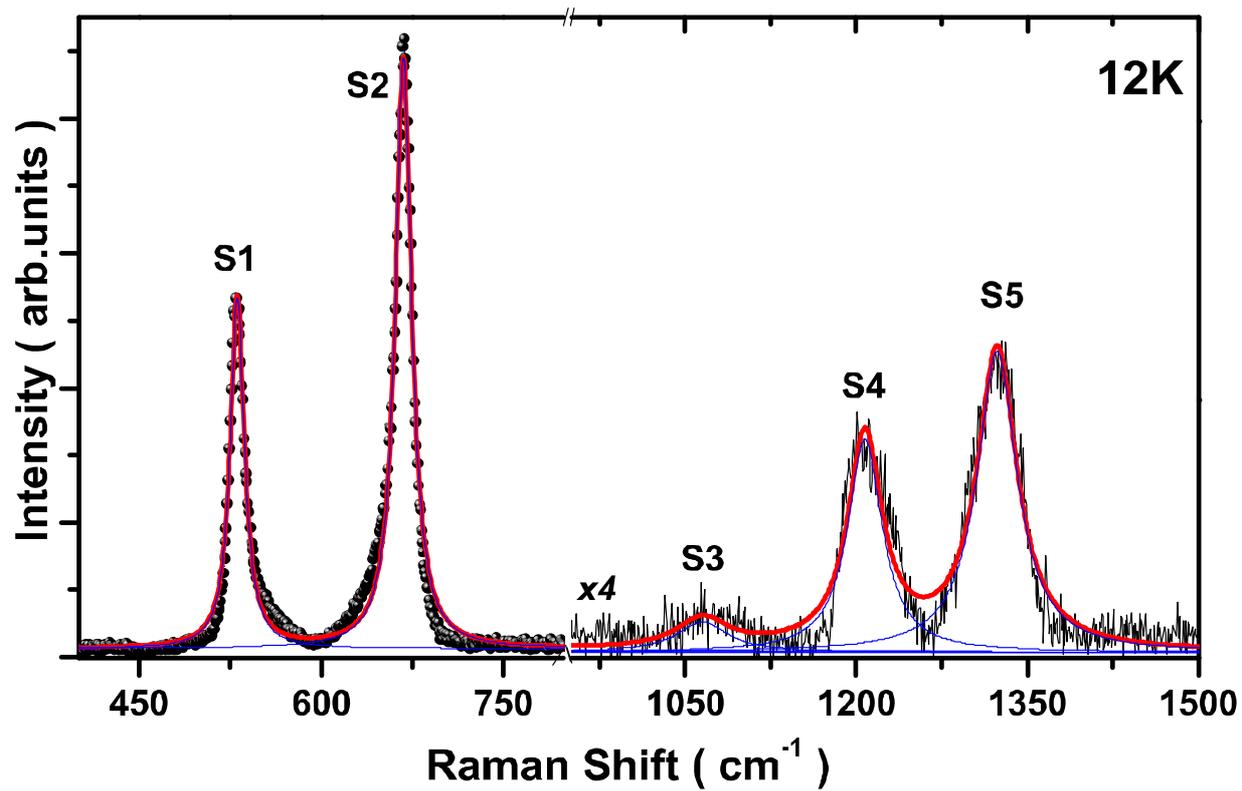

**Figure 6:**

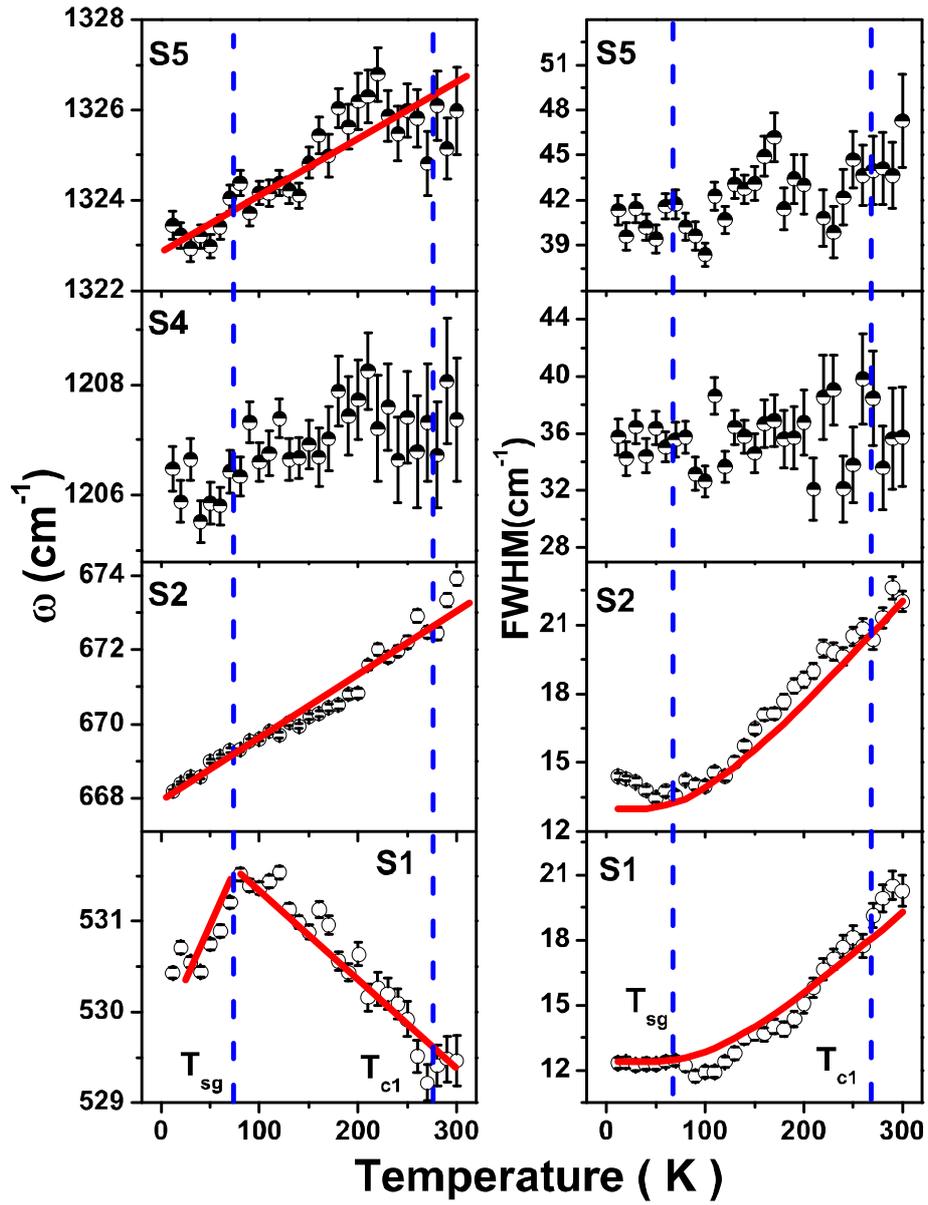

**Figure 7:**

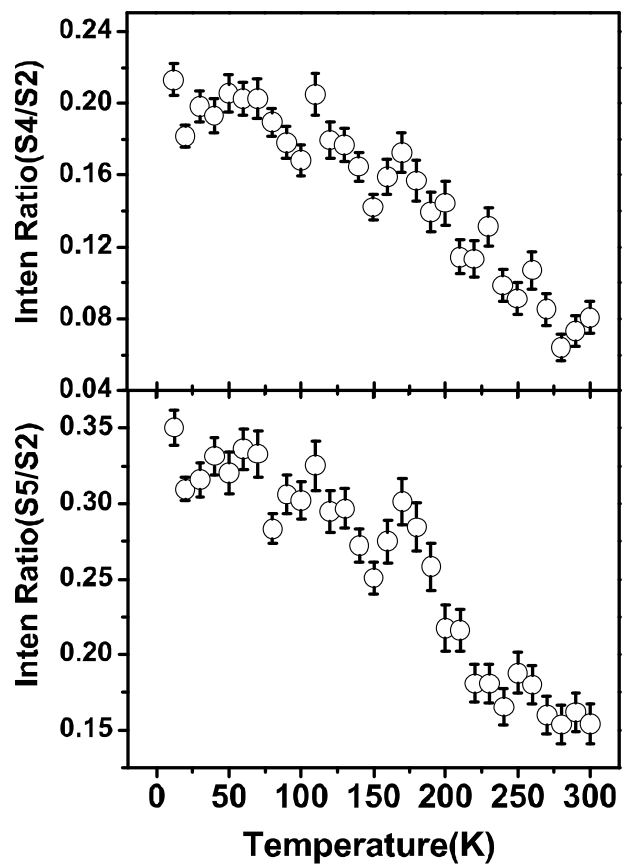